\documentclass[aps,prb,reprint,nofootinbib,twocolumn,superscriptaddress,showpacs,showkeys,longbibliography,amsmath,amssymb]{revtex4-2}
\usepackage{graphicx}
\usepackage{dcolumn}
\usepackage{bm}
\usepackage{braket}
\usepackage{subfigure}
\usepackage[colorlinks,bookmarks=false,citecolor=blue,linkcolor=red,urlcolor=blue]{hyperref}
\usepackage[english]{babel}
\usepackage{times} 
\usepackage{helvet}
\usepackage{courier}

\begin{document}
\preprint{APS/123-QED}
\title{Impact of Nonreciprocal Hopping on Localization in Non-Hermitian Quasiperiodic Systems}

\author{Xianqi Tong}\email[Corresponding author. Email: ] {1000006854@ujs.edu.cn}
\affiliation{Department of Physics, Jiangsu University, Zhenjiang 212013, China}
\author{Yiling Zhang}
\affiliation{School of Physical Science and Technology, and Collaborative Innovation Center of Suzhou Nano Science and Technology, Soochow University, 1 Shizi Street, Suzhou 215006, China}
\author{Bin Li}
\affiliation{Department of Physics, Zhejiang Normal University, Jinhua 321004, People's Republic of China}
\author{Xiaosen Yang}\email[Corresponding author. Email: ] {yangxs@ujs.edu.cn}
\affiliation{Department of Physics, Jiangsu University, Zhenjiang 212013, China}

\date{\today}

\begin{abstract}

We study the non-Hermitian Aubry-André-Harper model, incorporating complex phase modulation, unmodulated and modulated nonreciprocal hopping. Using Avila's global theory, we derive analytical phase boundaries and map out the phase diagrams, revealing extended, localized, critical, and skin phases unique to non-Hermitian systems. For complex phase modulation, we determine localization lengths through Lyapunov exponents and show that topological transitions align with localization transitions. In the nonreciprocal case, we use similarity transformations to confirm phase boundaries consistent with Avila's theory and uncover asymmetric localization behaviors. Importantly, modulated nonreciprocal hopping transforms both extended and critical phases into skin phases under open boundary conditions. These results highlight the interplay between topology, localization, and non-Hermitian effects, offering new perspectives on quasiperiodic systems.

\end{abstract}

\maketitle

\section{Introduction}
Anderson localization is the most fundamental research area in condensed matter physics \cite{Anderson1}. This concept play a significant role in many different area of physics, such as disordered photonics, ultracold atomic, condensed matter physics \cite{Anderson_cmp1, Anderson_cmp2, Anderson_cmp3, Anderson_cmp4, Anderson_ua2, Anderson_ua3, Anderson_ua4, Anderson_ua5, Anderson_ua6, Anderson_ua7, Anderson_ua8, Anderson_ua9, Anderson_photonic1, Anderson_photonic2, Anderson_photonic3, Anderson_photonic4, Anderson_photonic5, Anderson_photonic6, Anderson_photonic7}. Anderson predicted that when the system with uncorrelated potential in one-dimensional and two-dimensional case, all states are localized in random positions, while the extended-localized phase transition will happened in three-dimensional system when cross a disorder threshold \cite{Anderson2}. However, the disorder threshold can also arise in certain one-dimensional system with incommensurate potential or correlated disorder \cite{Anderson_ip1, Anderson_ip2, Anderson_ip3, Anderson_ip4, Anderson_ip5, Anderson_ip6}. The extended and localized state can coexist in the same spectrum and separated by a certain threshold termed as mobility edge \cite{Anderson_mobility1, Anderson_mobility2}. Many-body localization has been found in the one-dimension system with the interplay between the particle interaction and the uncorrelated disorder or incommensurate disordered system\cite{many-body_localization1, many-body_localization2, many-body_localization3}.

Quasiperiodic systems, represented by the celebrated Aubry-André-Harper (AAH) model, have long provided a fertile ground for studying localization phenomena. Unlike fully disordered systems, where Anderson localization occurs universally in one dimension, the AAH model exhibits a distinct transition between extended and localized phases at a critical modulation strength \cite{Anderson_ip1}. This self-duality relation \cite{AAH3}, rooted in the interplay of lattice structure and quasiperiodicity, connects the AAH model to topological physics, including Thouless pumping \cite{AAH7} and the Hofstadter butterfly \cite{AAH4}. Extensions of the AAH model, incorporating interactions, long-range couplings, or additional symmetries, have revealed rich physics, including mobility edges and multifractality \cite{Anderson_ip6, ME1,ME2, ME3, ME4, ME5,ME6,ME7,ME8,ME9,ME10,ME11,ME12,ME13,ME14}. Further research uncover that the whole critical phase exists in the AAH model with $p$-wave superconductivity or off-diagonal hopping \cite{pwave_1, pwave_2,pwave_3, offdiagonal0, offdiagonal1, offdiagonal2, offdiagonal3}.

The study of non-Hermitian systems has become a rapidly growing frontier in condensed matter physics, uncovering novel phenomena without Hermitian counterparts. These systems, characterized by the presence of gain, loss, or nonreciprocal hopping terms, exhibit striking features such as the non-Hermitian skin effect (NHSE) \cite{NHSE1,NHSE2,NHSE3,NHSE4,NHSE5,NHSE6,NHSE7,NHSE8,NHSE9}, parity-time (\(\mathcal{PT}\)) symmetry breaking\cite{PT1,PT2,PT3,PT4,PT5}, and exceptional points\cite{EP1,EP2}. Of particular interest is the breakdown of the conventional bulk-boundary correspondence, necessitating new approaches to understand topological properties in non-Hermitian regimes \cite{non-Hermitian7, non-Hermitian11, non-Bloch_BBC1}. Such explorations not only deepen our understanding of fundamental quantum mechanics but also hold promise for practical applications in photonics, acoustics, and electrical circuits \cite{RLC1,RLC2,RLC3,RLC4,RLC5,RLC6,acoustic1,acoustic2, photonic1, photonic2}.

Introducing non-Hermitian elements into the AAH model significantly enriches its phenomenology by unveiling novel physical effects absent in Hermitian systems \cite{AAH1,NHAAH1,NHAAH2,NHAAH3,PT1,PT2,PT3,PT4,PT5}. Nonreciprocal hopping terms, for example, induce the NHSE under open boundary condition (OBC), resulting in an accumulation of eigenstates at one edge of the system. Such an interplay between quasiperiodicity, localization, and NHSE has garnered significant interest in recent years, as it provides a platform for exploring the coexistence and competition of different non-Hermitian phenomena. The nonreciprocal AAH model not only retains its localization transition but also exhibits asymmetric localization lengths, dual Lyapunov exponents (LEs), and boundary-sensitive spectra \cite{NHAAH1,NHAAH2,NHAAH3}. Moreover, the topological nature of the localization transition, as characterized by winding numbers of complex eigenenergies under periodic boundary condition (PBC), establishes a bulk-bulk correspondence. However, an open question remains: how does the competition between unmodulated and modulated nonreciprocal hopping terms affect the phase diagram? More specifically, will the critical phase, like the extended and localized phases, also be influenced by non-Hermiticity in such systems?

In this work, we present a detailed study of the non-Hermitian AAH model, considering three key non-Hermitian ingredients: complex phase modulation, nonreciprocal unmodulated hopping, and nonreciprocal modulated hopping. By employing both numerical simulations and analytical approaches such as Avila's global theory, we comprehensively map out the phase diagram and derive exact analytical expressions for the phase boundaries. Our analysis explores \(\mathcal{PT}\) symmetry breaking, spectral topology, and localization transitions in the presence of complex potentials. Furthermore, we investigate the asymmetric localization behavior induced by nonreciprocal hopping through similarity transformations, revealing novel skin and critical phases. Notably, the introduction of modulated nonreciprocal hopping enables a unique transformation of critical phase into skin phase, which we characterize through Lyapunov exponents and spectral winding numbers. These findings provide significant insights into the interplay of topology, localization, and non-Hermitian effects in quasiperiodic systems.

The remainder of this paper is organized as follows. In Sec. \ref{Model and Hamiltonian}, we introduce the non-Hermitian AAH model. In Sec. \ref{localization and topology with the non-Hermitian AAH model}, we employ Avila's global theory to analyze the Lyapunov exponents and define topological invariants in the presence of complex phases and nonreciprocal hopping. Sec. \ref{localization and topology with the complex potential}, we examine the localization transitions, Lyapunov exponents, spectral topology, and \(\mathcal{PT}\) symmetry breaking for systems with complex phase modulation. Sec. \ref{ASYMMETRICAL LOCALIZATION IN THE PRESENCE OF NONRECIPROCAL HOPPING}, is divided into two parts: first, we analyze the unmodulated nonreciprocal hopping case, highlighting the interplay between localization and non-Hermitian effects; second, we study the modulated nonreciprocal hopping case, demonstrating how it transforms critical phase into skin phase. Finally, in Sec. \ref{CONCLUSION AND DISCUSSION}, we summarize our findings and discuss potential extensions to other quasiperiodic systems.

\section{Model and Hamiltonian}
\label{Model and Hamiltonian}
In this work, we study a tight-binding model with quasiperiodic modulations, described by the Hamiltonian:
\begin{equation} 
H = \sum_j \left[ t_j^L c_j^{\dagger} c_{j+1} + t_j^R c_{j+1}^{\dagger} c_j + V_j c_j^{\dagger} c_j \right], 
\label{eq_NH_Hamiltonian}
\end{equation}
where $c_j^{\dagger}$ ($c_j$) are the creation (annihilation) operators at site $j$. The hopping amplitudes $t_j^L$ and $t_j^R$ introduce nonreciprocal hopping leading to the NHSE \cite{NHSE1}. Meanwhile, $V_j$ represents the modulated on-site complex potential. The parameters are defined as:
\begin{equation}
\begin{aligned}
t_j^L & = t_L+\mu_L \cos [2 \pi \alpha(j+1 / 2)+\theta], \\
t_j^R& = t_R+\mu_R \cos [2 \pi \alpha(j+1 / 2)+\theta], \\
V_j & =V \exp [ - i (2 \pi \alpha j + i \phi + \theta) ].
\end{aligned}
\label{eq_tL_tR_Vj}
\end{equation}
Here, $t_{L/R}$ are the unmodulated components of the nonreciprocal hopping amplitudes, while $\mu_{L/R}$ and $V$ denote the modulation amplitudes of the off-diagonal hopping and on-site potential, respectively. The phase $\theta$ governs the relative phase between the modulations, and $\phi$ introduces an imaginary component into the potential. The quasiperiodicity of the modulations is determined by the irrational number $\alpha$, typically chosen as the inverse of the golden ratio, $\beta = (\sqrt{5} - 1) / 2$. Numerically, $\alpha$ is approximated by the ratio of consecutive Fibonacci numbers, $\alpha = F_n / F_{n+1}$, ensuring a commensurate lattice structure with $L = F_{n+1}$ sites.

\section{localization and topology with the non-Hermitian AAH model}
\label{localization and topology with the non-Hermitian AAH model}
To investigate localization phase transition under the presence of a complex potential, we analytically compute the LE, which corresponds to the inverse of the localization length, $\gamma=1/\xi$, for single-particle eigenstates. The behavior of a single-particle state \( |\psi\rangle = \sum_j \phi_j c^\dagger_j |0\rangle \), with eigenenergy \( E \), is described by the Schrödinger equation, which can be expressed in the form of a transfer matrix:
\begin{equation}
\left[\begin{array}{c}
\phi_{j+1} \\
\phi_j
\end{array}\right]=T_j\left[\begin{array}{c}
\phi_j \\
\phi_{j-1}
\end{array}\right], \quad T_j=\left[\begin{array}{cc}
\frac{E-V_j}{t_j^R} & -\frac{t_{j-1}^L}{t_j^R} \\
1 & 0
\end{array}\right].
\label{eq_transfer_matrix_potential}
\end{equation}
The LE of the state is computed by
\begin{equation}
\gamma_{\varepsilon}=\lim _{L \rightarrow \infty} \frac{1}{L} \ln \left\|\prod_{j=1}^L T_j(\theta+i \varepsilon)\right\|,
\end{equation}
where $ \|\cdot\| $ denoted the norm of the matrix. Notice that the analytical continuation of the global phase $(\theta \rightarrow \theta + i \varepsilon)$ has been performed, which plays a crucial role in Avila's global theory. By factorizing out the unbounded term $(t_j^R)^{-1}$ in the matrix $T_j$, we can rewrite the transfer matrix as a commutative product. Then, the LE for the unbounded part can be obtained by turning the product into a summation and then into an integral. Regarding the LE for the remaining matrix, Avila proved that for transfer matrices in the family of one-frequency analytical SL$(2, \mathbb{C})$ cocycles, the LE $\gamma_{\varepsilon=0}$ can be obtained from $\gamma_{\varepsilon \rightarrow \infty}$ (see. the Appendix \ref{COMPUTATION_OF_THE_LYAPUNOV_EXPONENT} for details) \cite{Avila1}. It yields two parts,
\begin{equation}
\gamma= \begin{cases}\max \left(f_1, 0\right), & \mu_R \leqslant t \\ \max \left(f_2, 0\right), & \mu_R>t\end{cases}
\label{eq_gamma}
\end{equation}
where
\begin{equation}
\begin{aligned}
& f_1=\max \left\{\ln \frac{\left| V e^{ \phi} \pm \sqrt{(Ve^{ \phi})^2-\mu_R\mu_L} \right|}{t_R+\sqrt{t_R^2-\mu_R^2}}\right\}, \\
& f_2=\max \left\{\ln \frac{\left| V e^{ \phi} \pm \sqrt{(Ve^{ \phi})^2-\mu_R\mu_L} \right|}{\mu_R}\right\} .
\end{aligned}
\label{eq_f1_f2}
\end{equation}
A state is classified as extended or critical if the LE $\gamma = 0$, and as localized if $\gamma > 0$. The transition from the extended or critical phase to the localized phase is determined by the conditions $f_1 = 0$ or $f_2 = 0$, where the LE shifts from $\gamma = 0$ to $\gamma > 0$.

For the condition $f_1 = 0$, the critical point is given by:
\begin{equation}
\begin{aligned}
V e^{\phi} = \frac{\mu_L \left(t_R-\sqrt{t_R^2-\mu _L^2}\right)+\mu _R \left(\sqrt{t_R^2-\mu _L^2}+t_R\right)}{2 \mu _R}\\
 \text { when } \quad  \mu_R \leqslant t_R \text {, }
\end{aligned}
\label{eq_transition_mutL} 
\end{equation}For $f_2 = 0$, the critical point is:
\begin{equation}
V e^{\phi}=\frac{\mu_L+\mu_R}{2} \quad \text{when} \quad \mu_R > t_R.
\label{eq_transition_tLmu}
\end{equation}

Notably, the condition $f_2 = 0$ spans a significant region of the parameter space, indicating the presence of a robust critical phase. The LE and the phase transition conditions are independent of the energy, meaning that no mobility edge is present. Additionally, these results are unaffected by the irrational frequency $\alpha$ or the global phase $\theta$. The localization properties in this regime remain similar to those in the Hermitian limit $(\phi = 0)$ \cite{Anderson_ip2}.

The imaginary phase $i\phi$ primarily renormalizes the potential strength $V$ by a factor of $e^{|\phi|}$. Apart from this renormalization, the localization properties remain consistent with those of the Hermitian case ($h = 0$). This demonstrates that introducing the imaginary phase modifies the system’s parameters without altering the fundamental nature of localization.

To further investigate the spectral properties of the system, we analyze the topology of the energy spectrum using the winding number. The winding number quantifies how the complex energy spectrum encircles a reference energy in the complex plane as the system parameter evolves. It is defined as \cite{NHAAH1, AAH1}:
\begin{equation}
 v_{E_B} = \lim_{L \to \infty} \frac{1}{2\pi i} \int_0^{2\pi / L} d\theta \frac{\partial}{\partial \theta} \ln \left[ \operatorname{det}(H - E_B) \right],
\label{eq_winding_number}
\end{equation}
where \(E_B\) is a chosen base energy, and \(\theta\) acts as an auxiliary parameter analogous to a momentum phase.

The winding number captures the topological behavior of the spectral trajectory as \(\theta\) evolves from 0 to \(2\pi\). A nonzero winding number indicates that the energy spectrum forms nontrivial loops around \(E_B\) in the complex energy plane \cite{NHSE1, NHSE4, NHAAH1, AAH1}. These loops reflect the intricate spectral structure arising from non-Hermitian effects. Notably, the imaginary phase \(i h\) renders the momentum phase \(\theta\) complex, which enriching the topological structure and resulting in nontrivial winding numbers. Different choices of \(E_B\) correspond to distinct loop structures within the spectrum. To focus on the most significant spectral features, we consider the most nontrivial winding number, given by $v = \operatorname{sgn}(v_{E_B}) \times \max\left(\left| v_{E_B} \right|\right), \forall E_B \in \mathbb{C}$. 
This approach identifies the presence of spectral loops in non-Hermitian energy spectra.

\section{localization and topology with the complex potential}
\label{localization and topology with the complex potential}

\begin{figure}[]
	\centering
	\includegraphics[width=3.4in]{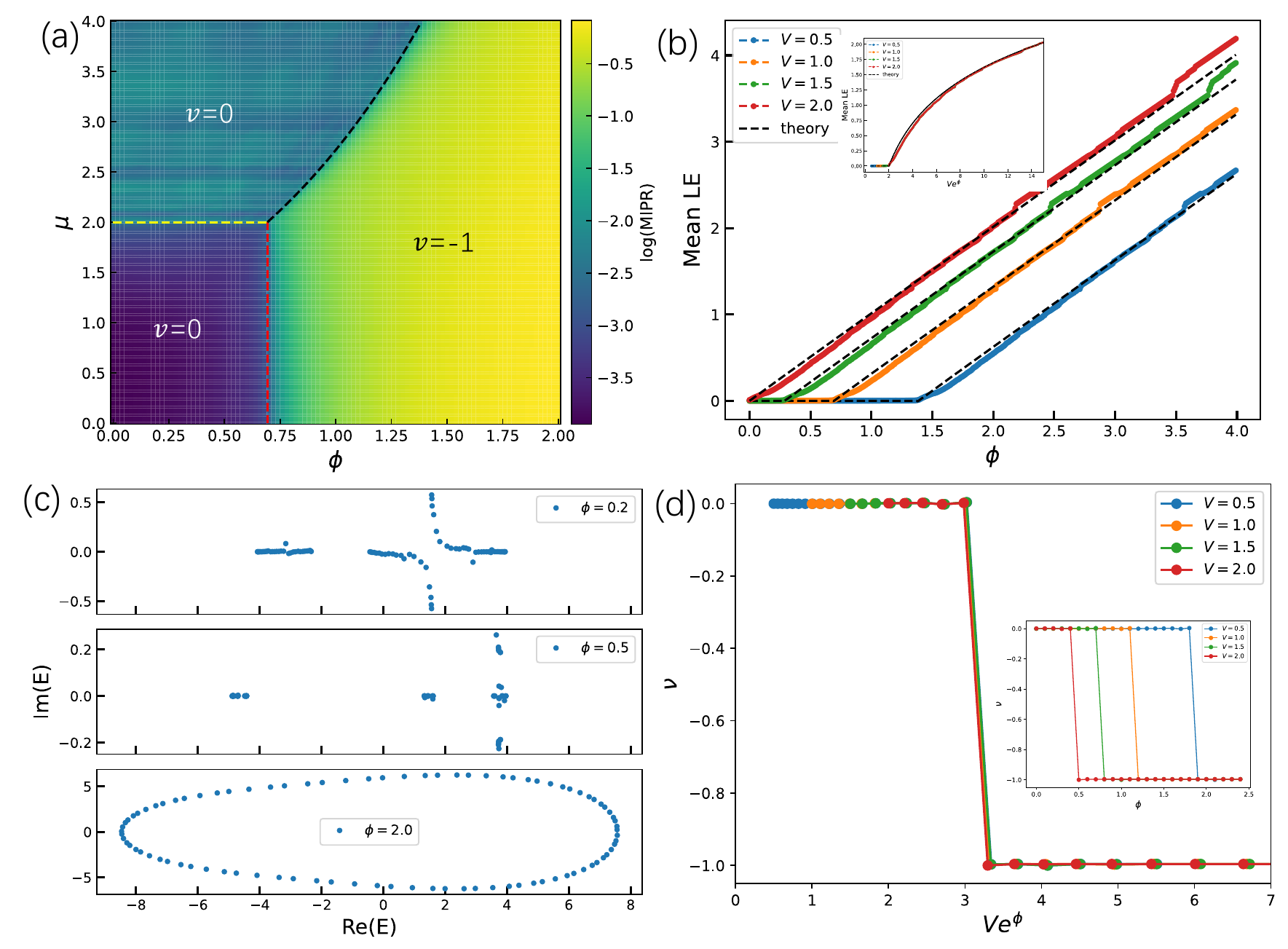}
	\caption{ In (a), $\log(\text{MIPR})$ in the $(\phi, \mu)$ plane for $V = 1$. 
(b) Mean LE as a function of $\phi$ for $V = 0.5, 1.0, 1.5, 2.0$, with analytical results shown by black-dotted lines. The inset of (b) shows the rescaled mean LE as a function of $V e^\phi$.
(c) Complex energy spectrum for $(\mu, \phi) = (0.5, 0.2)$, $(2.5, 0.5)$, and $(2.5, 2.0)$. 
In (d) and its inset show the Winding number $\nu$ versus rescaled phase $V e^\phi$ and unrescaled phase $\phi$ for $V = 0.5, 1.0, 1.5, 2.0$. The parameters are $L = 89$ and $\theta = 0$.}
\label{fig_paradiagram_phi}
\end{figure}
In the symmetric case $t_L = t_R = t$, $\mu_L = \mu_R = \mu$, the localization transition is governed by the condition, Eqs.$~$(\ref{eq_transition_mutL}) and (\ref{eq_transition_tLmu}):
\begin{equation}
    \phi = \ln \frac{\max (t, \mu)}{V},  
    \label{eq_boundary_complex_potential}
\end{equation}
which separates different quantum phases. When $\phi$ exceeds this critical value, the system enters a localized phase where all states exhibit a positive, energy-independent LEs ($\gamma > 0$). For $\phi < \ln(t/V)$ and $\mu < t$, we have $f_1 <0$ (Eq.~(\ref{eq_transition_mutL})), leading to a vanishing Lyapunov exponent ($\gamma = 0$) and an extended phase. Similarly, for $\phi < \ln(\mu/V)$ and $\mu > t$, $f_2 < 0$ (Eq.~(\ref{eq_transition_tLmu})) characterizes a critical phase, where all states are critical.

To validate these analytical predictions, we compute the mean inverse participation ratio (MIPR). For a normalized eigenstate, the MIPR is defined as: $\text{MIPR} = \frac{1}{L} \sum_{j,n=1}^L |\phi_j^n|^4$, where $\left| \phi_j^n \right|$ is the amplitude of $n$-th eigenvector at site $j$. In the extended phase, $\text{MIPR} \sim L^{-1}$, while in the localized phase, $\text{MIPR} \sim L^0$. In the critical phase, the MIPR takes intermediate values, $0 < \text{MIPR} < 1$, reflecting the multifractal nature of the eigenstates. 

The logarithm of MIPR, $\log\text{(MIPR)}$, in the $(\phi, V)$-plane, as shown in Fig.~\ref{fig_paradiagram_phi}(a), clearly delineates the phase boundaries, which agree well with the theoretical prediction. When both \( \phi \) and \( \mu \) are small, the system resides in the extended phase, characterized by  \( \log( \text{MIPR} )\sim -\log L\). For \( \mu > \phi \), \( -\log L < \log (\text{MIPR}) < 0 \), indicating that the system has transitioned to the critical phase. As \(V\) increases sufficiently, the system enters the localized phase, where the wavefunctions become insensitive to boundary conditions. The red and black dashed lines separate the extended-localized and critical-localized phases, respectively, with the yellow dashed-dotted line at $\mu=t$ dividing the extended and critical phases.

To further investigate the localization transition, we analyze the LE by fitting eigenstates to exponential wavefunctions:
\begin{equation}
    \phi^n_j \sim \exp(-\gamma_n |j - j_0|),
\end{equation}
where $j_0$ is the localization center.  The mean LE $\gamma = \sum_n \gamma_n/L$, averaged over all eigenstates, is presented in Fig.~\ref{fig_paradiagram_phi}(b) for three values of $V = 0.5, 1.0, 1.5, 2.0$. At the transition points predicted by Eq.~(\ref{eq_gamma}), the mean LE $\gamma$ rises sharply from zero, marking the onset of localization and and aligning with the analytical prediction, as represented by the black dashed line. In the panel of Fig.~\ref{fig_paradiagram_phi}(b), we perform a rescaling using $V e^{\phi}$, and observe that all the curves collapse onto a single line, which agrees well with the analytical solution \cite{offdiagonal2}.

We also investigate the spectral properties of the system as a function of $\phi$. Fig.~\ref{fig_paradiagram_phi}(c) presents the evolution of the complex energy spectrum as $\phi$ increases. A \(\mathcal{PT}\)-symmetry transition is most prominent at $\phi = 0$, where the imaginary part of the eigenvalues $|\text{Im}(E)|$ shows no sharp increase for $\phi > 0$ \cite{AAH1}. In the localized phase, loops emerge in the spectrum, whereas these loops are absent in the extended and critical phases. This observation is quantified by the winding number of the spectral trajectory in the complex plane using Eq.~(\ref{eq_winding_number}), as shown in Fig.~\ref{fig_paradiagram_phi}(d). The winding number for $V =0.5, 1.0, 1.5, 2.0$ are plotted after rescaling, and all the curves collapse onto a single line, highlighting its universality across different parameter regimes. The localization transition points, indicated by the dotted lines, coincide perfectly with the topological phase transitions. In the subplot, the winding numbers without rescaling are shown.

\section{ASYMMETRICAL LOCALIZATION IN THE PRESENCE OF NONRECIPROCAL HOPPING}
\label{ASYMMETRICAL LOCALIZATION IN THE PRESENCE OF NONRECIPROCAL HOPPING}

\subsection{Case $t_R \neq t_L$}
\begin{figure}[]
	\centering
	\includegraphics[width=3.4in]{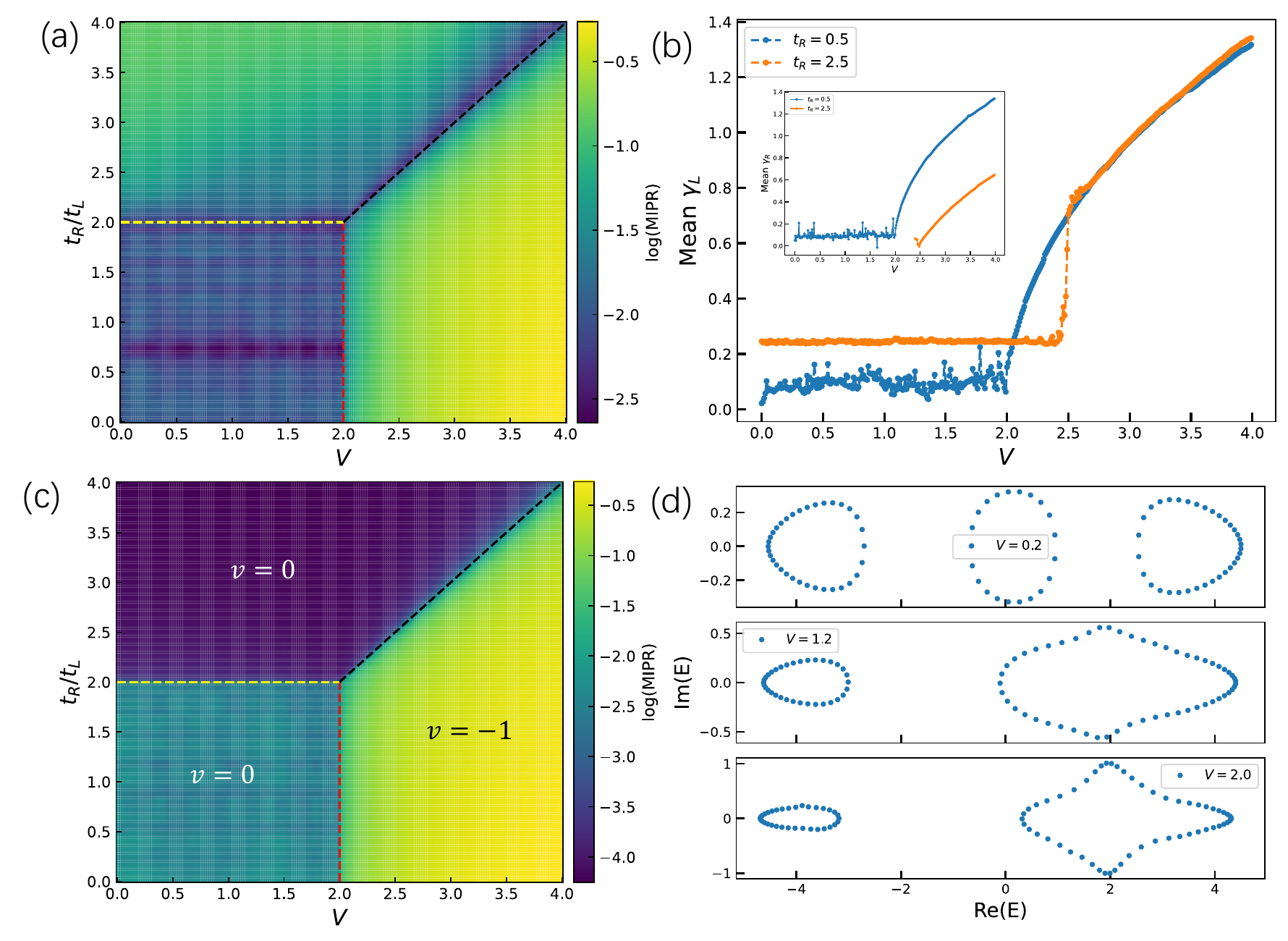}
	\caption{In (a), $\log(\text{MIPR})$ in the $(V, t_R)$ plane under OBC. 
(b) Mean left side LEs $\gamma_L$ as a function of $V$ for $t_R = 0.5, 2.5$. The inset of (b) shows $\gamma_R$ as a function of $V$. 
(c) $\log(\text{MIPR})$ in the $(V, t_R)$ plane under PBC with $\phi = 0$. 
(d) Complex energy spectrum for $(V, t_L) = (0.2, 2.5)$, $(1.2, 2.5)$, and $(2.0, 2.5)$. The parameters are $L = 89$, $\mu = 1$, $\phi = 0$, and $\theta = 0$.}
\label{fig_paradiagram_tc}
\end{figure}
Nonreciprocal hopping, where $t_R \neq t_L$, gives rise to NHSE, manifesting as a boundary-dependent spectrum and eigenstates. To understand this phenomenon, we first focus on the case under OBC. When $\mu_R=\mu_L=\mu$, Eq.~(\ref{eq_f1_f2}) simplifies to the following form:
\begin{equation}
\begin{aligned}
& f_1=\max \left\{\ln \frac{\left| V e^{\phi} \pm \sqrt{(Ve^{\phi})^2-\mu^2} \right|}{t_R+\sqrt{t_R^2-\mu^2}}\right\}, \\
& f_2=\max \left\{\ln \frac{\left| V e^{\phi} \pm \sqrt{(Ve^{\phi})^2-\mu^2} \right|}{\mu}\right\} .
\end{aligned}
\label{eq_f1_f2}
\end{equation}
Localization phase transition points are identified by solving $f_{1(2)} = 0$. Specifically:
\begin{equation}
V e^{\left| \phi \right|} = \mu \quad \text{when} \quad  t_R < \mu,
\label{eq_transition_tc1}
\end{equation} 
and
\begin{equation}
V e^ {\left| \phi \right|} = t_R \quad  \text{when}  \quad t_R \geq \mu.
\label{eq_transition_tc2}
\end{equation} 

For small $V$, the system is in the skin phase or the critical phase, as illustrated in Fig.~\ref{fig_paradiagram_tc}. In the skin phase, all eigenstates localize asymmetrically on one edge. Specifically, when $t_R ( t_L) >\mu$, states are localized at the right $(\text{or left})$ edge depending on the strength of the nonreciprocity. Conversely, when $t_R<\mu$, eigenstates remain critical, and it is worth noting that the critical phase is not affected by nonreciprocal hopping, which contrasts with conventional non-Hermitian systems \cite{NHAAH2}. For sufficiently large $V$, the system transitions to the localized phase, where all eigenstates are fully localized. For simplicity, the following discussion focuses on $t_R$ with $\phi = 0$.

The wavefunction in the skin phase is analyzed using a similarity transformation. The Hamiltonian in Eq.~(\ref{eq_NH_Hamiltonian}) is mapped to
\begin{equation}
H^{'}=\sum_j \left[ t_j^{'} b_j^{\dagger} b_{j+1}+t_j^{'} b_{j+1}^{\dagger} b_j+V_j b_j^{\dagger} b_j\right],
\end{equation}
where the transformation $c_j = dr_j b_j$ and $c_j^{\dagger} = dr_j^{-1} b_j^{\dagger}$ introduces a factor $dr_j = \prod_{i=1}^j \sqrt{\frac{t_j^R}{t_j^L}} = \prod_{i=1}^j r_j$, with $r_j = \sqrt{\frac{t_j^R}{t_j^L}}$. In the reciprocal limit $( t_i^R = t_i^L )$, we find that $dr_j = 0$, and the system reduces to the Hamiltonian $H$ with only an imaginary potential $(\eta = 0)$. Consequently, the Hamiltonian $H^{'}$ exhibits the same single-particle physics as $H$, where nonreciprocal hopping vanishes. This transformation ensures that the spectra and winding numbers of $H^{'}$ and $H$ are identical under OBC.

For the wave functions $\phi_j^n$ of $H^{\prime}$, the corresponding wave functions of $H$ are related by the transformation $e^{\log dr_j} \phi_j^n$. The relative nonreciprocal hopping strength $t_R / t_L$ has a significant impact on the eigenstate amplitudes, particularly in the extended phase. When $t_R / t_L > \mu$, all the eigenstates become localized at the right boundary, with left side LEs satisfying $\gamma_L = \left| r_j \right| $. Conversely, when $t_R / t_L < \mu$, the off-diagonal hopping modulation dominates, driving the system into a critical phase. This phase remains unaffected by the nonreciprocal effects and does not transition into a skin effect phase.

In the localized phase, the wave function \( \phi_j^n \) is expressed as:
\begin{equation}
\begin{aligned}
\varphi_j^n \propto
\begin{cases}
e^{-(\gamma - \frac{1}{L} \log dr_j)(j - j_0)}, & \text{for } j > j_0, \\
e^{-(\gamma + \frac{1}{L} \log dr_j)(j_0 - j)}, & \text{for } j < j_0,
\end{cases}
\end{aligned}
\label{eq_transformation_localizaiton}
\end{equation}
where \( \gamma \) is defined by Eq.~(8) for \( t_j^R = t_j^L \). In this phase, the right-side and left-side LEs are given as \( \gamma_R = \gamma - \frac{1}{L} \log dr_j \) and \( \gamma_L = \gamma + \frac{1}{L} \log dr_j \), respectively. Consequently, nonreciprocal hopping (\( \log dr_j \neq 0 \)) leads to asymmetrical localization of the wave function. When \( \gamma - | \frac{1}{L} \log dr_j| > 0 \), the wave function is localized at a position \( j_0 \), whereas for \( \gamma - | \frac{1}{L} \log dr_j| < 0 \), all bulk states localize at a boundary, resulting in the skin phase. The localization phase transition points are determined by the condition \( \gamma - | \frac{1}{L} \log dr_j| = 0 \), which is consistent with Eq.~(\ref{eq_f1_f2}). Detailed derivations can be found in Appendix~\ref{sec_similarity_transformation}.

In Fig.~\ref{fig_paradiagram_tc}(a), we present the log(MIPR) as a function of \(V\) and \(t_c\) under OBC. For \(t_R < \mu\), the system is in the critical phase with wavefunctions exhibiting fractal properties, characterized by \(\gamma = 0\). When \(t_R > \mu\), the wavefunctions localize at the right boundary. In both the localized phase at large \(V\) and the skin phase for \(t_R > \mu\), \(\log(\text{MIPR})\) approaches unity. In the skin phase, \(\gamma^L = r_j\), while the right Lyapunov exponent \(\gamma^R\) is undefined. In the localized phase, the wavefunctions are asymmetrically localized, with \(\gamma^L = \gamma^R + 2r_j\), reflecting the distinct localization properties at the left and right boundaries.

Although both the skin and extended phases exhibit localized wavefunctions, their localization mechanisms differ fundamentally. In Fig.~\ref{fig_paradiagram_tc}(b), we present the average \(\gamma_L\) for the different phases. The presence of nonreciprocal hopping leads to finite \(\gamma^L\) in both the skin and critical phases. As \(V\) increases, \(\gamma^L\) undergoes a discontinuous jump of \(2r_j\), highlighting the distinct localization behavior induced by the asymmetry in hopping amplitudes. Moreover, in the subplot of Fig.~\ref{fig_paradiagram_tc}(b), \(\gamma_R\) remains undefined in the skin phase because all eigenstates are localized at the right edge.

Under PBC, the similarity transformation no longer holds, resulting in distinct characteristics for each phase in terms of eigenstates and energy spectra. We compute the complex energy spectrum for various phases under PBC, as shown in Fig.~\ref{fig_paradiagram_tc}(c). Notably, loops appear in the complex spectrum even in the extended and critical phases. However, the winding number alone cannot directly characterize these loops. Fig.~\ref{fig_paradiagram_tc}(d) illustrates the winding numbers across different phases, indicating that the topological phase transition points align with the localization transition points, as described by Eqs.~(\ref{eq_transition_tc1}) and (\ref{eq_transition_tc2}).  In the extended and critical phases, the winding number is zero, while it is -1 in the localized phase. This highlights that the topological properties in these phases are influenced by nonreciprocal hopping, which cannot be fully captured by the winding number $\nu$.

\subsection{Case $\mu_R \neq \mu_L$}
\begin{figure}[]
	\centering
	\includegraphics[width=3.4in]{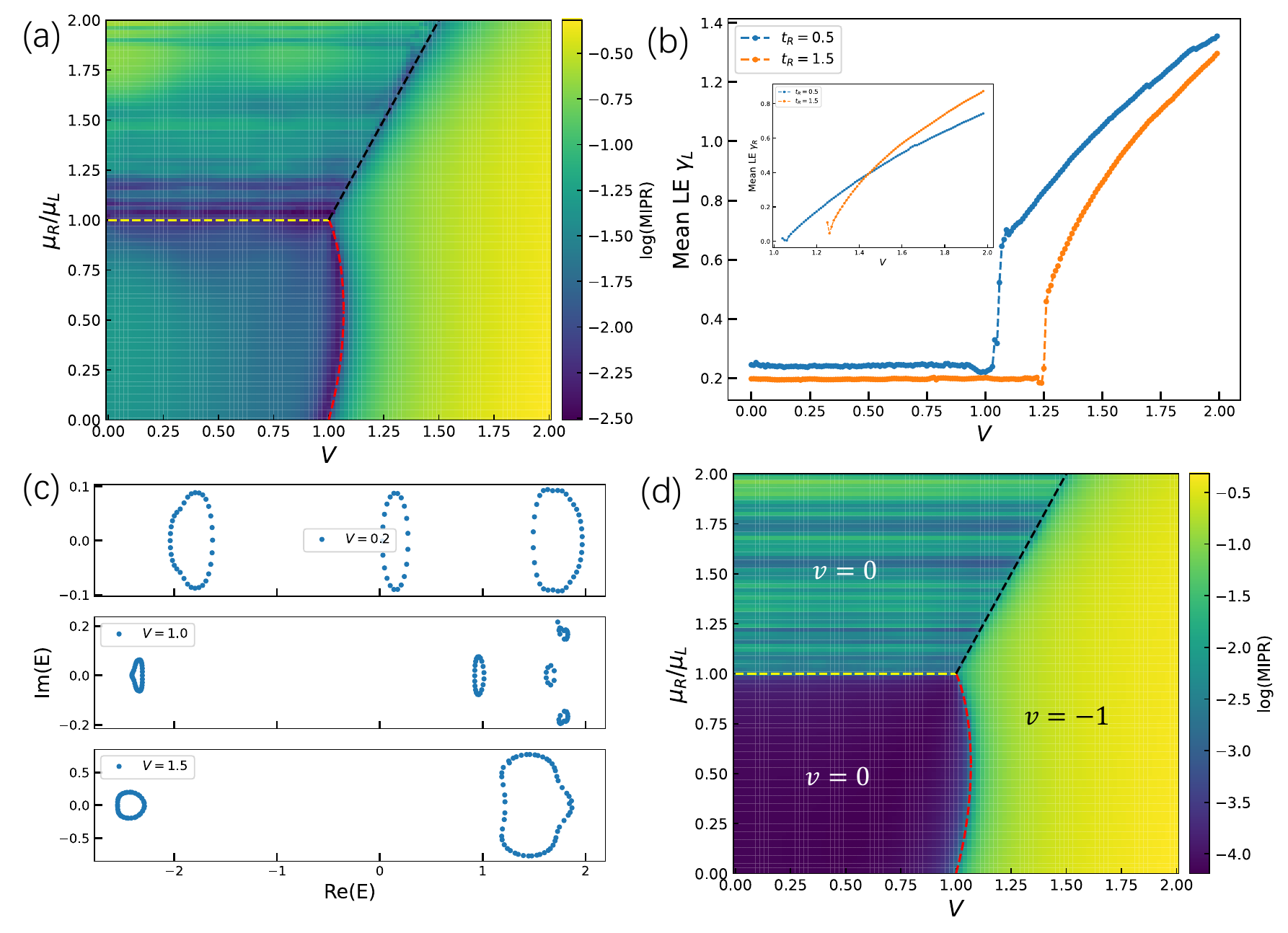}
	\caption{In (a) $\log(\text{MIPR})$ in the $(V, \mu_R)$ plane under OBC. 
(b) Left and right side LEs as a function of $V$ for $\mu_R = 0.5, 1.5$, with the inset showing the right side LE. 
(c) Complex energy spectrum for $(\mu, \phi) = (0.2, 0.5)$, $(1.0, 1.5)$, and $(1.5, 1.5)$. 
In (d), $\log(\text{MIPR})$ in the $(V, \mu_R)$ plane under PBC. The parameters are $L = 89$, $t = 1$, $\phi = 0$, and $\theta = 0$.}
\label{fig_paradiagram_mu}
\end{figure}
For $\mu_R \neq \mu_L$, $t_R = t_L = t$, the phase transition conditions are determined by Eqs.~(\ref{eq_transition_mutL}) and (\ref{eq_transition_tLmu}). Unlike the case of $t_R \neq t_L$, where only the extended phase transitions to the skin phase, the asymmetry between $\mu_R$ and $\mu_L$ induces a skin effect in both the extended and critical phases. This results in qualitatively different critical properties and a significantly altered phase diagram compared to the symmetric case.

Fig.~\ref{fig_paradiagram_mu}(a) shows $\log(\text{MIPR})$ as a function of $V$ and $\mu$ under OBC. For $\mu_R \neq \mu_L$, the nonreciprocal hopping modulation drives both the extended and critical phases into skin-localized states, leading to $\log(\text{MIPR}) \to 0$ across all three phases. This behavior contrasts with the $t_R \neq t_L$ case, where only the extended phase transitions into the skin phase while the critical phase remains unaffected. Thus, the asymmetry between $\mu_R$ and $\mu_L$ introduces a novel mechanism, transforming the critical phase into a skin phase and significantly altering the phase diagram. At the critical point, the wavefunctions are maximally extended, resulting in the minimal value of $\log(\text{MIPR})$, consistent with the theoretical prediction in Eq.~(\ref{eq_transition_mutL}).

The behavior of $\gamma_L$ across the phase diagram is analyzed in Fig.~\ref{fig_paradiagram_mu}(b). For $\mu_R \neq \mu_L$, both the extended and critical phases exhibit skin localization at the right boundary, leading to finite left side LEs. When $V$ is small, the system resides in the skin phases, and the right side $\gamma_R$ is undefined, as shown in the inset of Fig.~\ref{fig_paradiagram_mu}(b). At the phase transition points, the right side LEs sharply drop to values near zero, corresponding to the maximally extended states. As $V$ increases further, the right-side LEs grow again, indicating the onset of localization.

In Fig.~\ref{fig_paradiagram_mu}(c), we depict the complex energy spectra across different phases under PBC. The presence of a nonzero $\mu_R$ results in complex eigenenergies with winding loops observed in all three phases: extended, critical, and localized. This indicates that the nonreciprocal modulation not only alters the spatial properties of the eigenstates but also significantly impacts the spectral topology.

Winding numbers for different phases in the phase diagram are shown in Fig.~\ref{fig_paradiagram_mu}(d). Only the localized phase exhibits $\nu = -1$, while the winding number is zero in the other phases. In the PBC phase diagram, we observe that the skin phases evolve into extended and critical phases, with the topological phase transition aligning with the localization transition. Spectral loops persist across the phases, except at the transition points, and the spectra are complex when $\eta$, $h$, and $V$ are finite. Since the bulk states in the localized phase are insensitive to boundaries, this results in identical localization transition points for both PBC and OBC.

\section{CONCLUSION AND DISCUSSION}
\label{CONCLUSION AND DISCUSSION}
In this work, we have conducted a detailed study of the non-Hermitian AAH model, focusing on the effects of complex phase modulation, unmodulated nonreciprocal hopping, and modulated nonreciprocal hopping. By combining numerical simulations with Avila's global theory, we derived exact analytical expressions for the phase boundaries and systematically constructed the phase diagram. The study unveils the interplay between non-Hermiticity and quasiperiodicity, revealing a range of distinct phases, including extended, localized, critical, and skin phases.

For systems with complex phase modulation, we investigated the localization transitions, $\mathcal{PT}$ symmetry breaking, and spectral topology. Using LEs, we quantified the localization lengths of eigenstates, establishing a direct correspondence between topological and localization transitions. These findings show that introducing a complex phase significantly alters both spectral properties and the localization behavior of the system, thereby deepening our understanding of the connection between topology and non-Hermitian effects.

In the nonreciprocal hopping case, we analyzed two scenarios: unmodulated and modulated nonreciprocal hopping. For unmodulated nonreciprocal hopping, we demonstrated that extended phases under PBC transition into skin phases under OBC due to the NHSE, with asymmetric localization lengths, as confirmed by similarity transformations. For modulated nonreciprocal hopping, we showed that both critical and extended phases evolve into skin phases under OBCs. Additionally, we examined the spectral topology and found that while spectral loops persist across phases, winding numbers alone fail to capture their full characteristics. Nevertheless, topological and localization transitions remain aligned, emphasizing the robustness of the underlying physics.

This work establishes a unified framework for understanding the intricate relationships between topology, localization, and non-Hermitian effects in quasiperiodic systems. Future research will explore higher dimensional systems, the impact of interactions and disorder, and experimental realizations in photonic, acoustic, and cold atom platforms. By combining analytical techniques with numerical simulations, this study provides valuable insights and paves the way for future investigations into non-Hermitian quasiperiodic systems.

\begin{acknowledgments}
This work is supported by Natural Science Foundation of Jiangsu Province (Grant No. BK20231320).
\end{acknowledgments}

\appendix
\section{COMPUTATION OF THE LYAPUNOV EXPONENT} 
\label{COMPUTATION_OF_THE_LYAPUNOV_EXPONENT}
We factor out the unbounded term and rewrite the transfer matrix from Eq. (\ref{eq_transfer_matrix_potential}) as
\begin{equation}
T_j=\left[\begin{array}{cc}
\frac{E-V_j}{t_j^R} & -\frac{t_{j-1}^L}{t_j^R} \\
1 & 0
\end{array}\right]=A_j B_j
\end{equation}
where
\begin{equation}
\begin{aligned}
A_j & =\frac{1}{t_R + \mu_R \cos [2 \pi \alpha(j+1 / 2) + \theta]} \\
B_j & =\left[\begin{array}{cc}
E-V_j & -t_{j-1}^L \\
t_j^R & 0
\end{array}\right]
\end{aligned}
\end{equation}
$t_j^R$, $t_{j-1}^L$, and $V_j$ refer to the modulated off-diagonal hopping and on-site complex potential, respectively, as defined in Eqs. (\ref{eq_tL_tR_Vj}). The LE of the single-particle state is computed by
\begin{equation}
\begin{aligned}
\gamma(E) & =\lim _{L \rightarrow \infty} \frac{1}{L} \ln \left\|\prod_{j=1}^L T_j\right\| \\
& =\lim _{L \rightarrow \infty} \frac{1}{L} \ln \left\|\prod_{j=1}^L A_j\right\|+\lim _{L \rightarrow \infty} \frac{1}{L} \ln \left\|\prod_{j=1}^L B_j\right\| \\
& =\gamma^A(E)+\gamma^B(E)
\end{aligned}
\end{equation}
where the norm \( \left\| \cdot \right\| \) of a matrix is defined as the largest absolute value of its eigenvalues. Using ergodic theory [90] and Jensen’s formula \cite{Jensen1}, we compute
\begin{equation}
\begin{aligned}
&\begin{aligned}
\gamma^A(E) & =\lim _{L \rightarrow \infty} \frac{1}{L} \ln \prod_{j=1}^L \frac{1}{|t_R + \mu_R \cos [2 \pi \alpha(j + 1 / 2) + \theta]|} \\
& = \frac{1}{2 \pi} \int_0^{2 \pi} \ln \frac{1}{|t_R + \mu_R \cos \theta|} d \theta \\
& = \begin{cases} \ln \frac{2}{t_R + \sqrt{t_R^2 - \mu_R^2}}, & \mu_R \leqslant t_R, \\
\ln \frac{2}{\mu_R}, & \mu_R > t_R. \end{cases}
\end{aligned}\\
\end{aligned}
\end{equation}
To compute \(\gamma^B(E)\), we apply Avila’s global theory. The first step is to analytically continue the global phase, i.e., \(\theta \to \theta + i\varepsilon\), and in the absence of ambiguity, we use the same notation for a quantity and its continuation. As \(\varepsilon \to +\infty\), a direct calculation yields
\begin{equation}
\begin{aligned}
&B_j(\varepsilon \to +\infty) = e^{-i 2 \pi \alpha j + \varepsilon} \left[\begin{array}{cc}
- V e^{\phi} & -\mu_R e^{i \pi \alpha} / 2 \\
\mu_L e^{-i \pi \alpha} / 2 & 0
\end{array}\right] + o(1),
\end{aligned}
\end{equation}
which results in
\begin{equation}
\gamma_{\varepsilon \to +\infty} = \varepsilon + \text{max} \left\{ \text{ln} \frac{\left| V e^\phi \pm \sqrt{(Ve^\phi)^2 - \mu_R \mu_L} \right|}{2} \right\}.
\end{equation}
Based on Avila's global theory, \(\gamma_{\varepsilon}^B(E)\) is a convex, piecewise linear function with integer slopes. The energy \(E\) lies outside the spectrum if and only if \(\gamma_{\varepsilon=0}^B(E) > 0\). Additionally, $\gamma_{\varepsilon}^B(E)$ is an affine function in the neighborhood of $\varepsilon=0$.. Including \(\gamma^A(E)\), we obtain the LE of the single-particle state:
\begin{equation}
\gamma(E) = \begin{cases} \max \left( f + \ln \frac{2}{t_R + \sqrt{t_R^2 - \mu_R^2}}, 0 \right), & \mu_R \leqslant t_R, \\
\max \left( f + \ln \frac{2}{\mu_R}, 0 \right), & \mu_R > t_R \end{cases},
\end{equation}
where
\begin{equation}
f = \max \left\{ \begin{array}{l}
\ln \frac{ \left| V e^\phi + \sqrt{(V e^\phi)^2 - \mu_R \mu_L} \right| }{2} \\
\ln \frac{ \left| V e^\phi - \sqrt{(V e^\phi)^2 - \mu_R \mu_L} \right| }{2}
\end{array} \right\}.
\end{equation}

\section{$C_N$}
\label{sec_similarity_transformation}
To study Anderson localization, we rewrite the Hamiltonian in a biorthogonal basis as
\begin{equation}
H^{'} = \sum_j \left[ t_j^{'} c_j^{\dagger} c_{j+1} + t_j^{'} c_{j+1}^{\dagger} c_j + V_j c_j^{\dagger} c_j \right],
\end{equation}
where $H = \sum_{nm} h_{nm} \ket{n}\bra{m}$ is equivalent to $H^{'} = \sum_{nm} h_{nm}^{'} \ket{\widetilde{m}_R}\bra{\widetilde{n}_L}$, with $\ket{\widetilde{m}_R} = \prod_{j=1}^{m} r_j^{-1} \ket{m}, \bra{\widetilde{m}_R} = \prod_{j=1}^{m} r_j \bra{m}$, where $r_j = \sqrt{\frac{t_j^R}{t_j^L}}$. This transformation also reveals that all eigenenergies of the Hamiltonian are real, because $H$ and $H^{'}$ are similar with the relation $H^{'} = S^{-1} H S$, where $S = \text{diag}\left\{ \prod_{j=1}^{1} r_j, \prod_{j=1}^{2} r_j, \dots, \prod_{j=1}^{L} r_j \right\}$ is a diagonal matrix.
\begin{equation}
dr_j =  \prod_{i=1}^{L} r_j
\end{equation}
In the limit $N \to \infty$, we find that
\begin{equation}
\begin{aligned}
dr_j &=  \exp \left[ \sum_{i=1}^{L} \log \sqrt{ \frac{b_{i+1,i}}{b_{i,i+1}} } \right] \\
      &= \lim_{N \to \infty}  \exp \left[ \frac{L}{4 \pi} \int_0^{2\pi} d\theta \log \frac{ t_L + \mu_L \cos \theta }{t_R + \mu_R \cos \theta } \right].
\end{aligned}
\end{equation}
Thus, we get
\begin{equation}
\log dr_j = 
\begin{cases}
\frac{L}{2} \log \frac{\mu_L}{\mu_R}, & \mu_L > t_L \text{ and } \mu_R > t_R, \\
\frac{L}{2} \log \frac{ t_L + \sqrt{t_L^2 - \mu_L^2} }{\mu_R}, & \mu_L \leq t_L \text{ and } \mu_R > t_R, \\
\frac{L}{2} \log \frac{ \mu_L }{ t_R + \sqrt{t_R^2 - \mu_R^2} }, & \mu_L > t_L \text{ and } \mu_R \leq t_R, \\
\frac{L}{2} \log \frac{ t_L + \sqrt{t_L^2 - \mu_L^2} }{ t_R + \sqrt{t_R^2 - \mu_R^2} }, & \mu_L \leq t_L \text{ and } \mu_R \leq t_R.
\end{cases}
\label{eq_logdfj}
\end{equation}
For simplicity, we focus on the case $\mu_L = t_L$ in the main text.

By substituting Eq. (\ref{eq_logdfj}) into Eq. (\ref{eq_transformation_localizaiton}) in the main text and using $\gamma - | \frac{1}{L} \log dr_j| = 0$, we obtain the following expressions for $\mu_L \leq t_L$:
\begin{equation}
\begin{aligned}
 & f_1 = \frac{t_{R} + \sqrt{t_{R}^{2} - \mu_{R}^{2}}}{t_{L} + \sqrt{t_{L}^{2} - \mu_{L}^{2}}}, & \mu_{R} \leq t_{R} \\
 & f_2 = \frac{\mu_{R}}{t_{L} + \sqrt{t_{L}^{2} - \mu_{L}^{2}}}, & \mu_{R} > t_{R}
\end{aligned}
\end{equation}
For $\mu_L > t_L$:
\begin{equation}
\begin{aligned}
 & f_1 = \frac{t_{R} + \sqrt{t_{R}^{2} - \mu_{R}^{2}}}{\mu_{L}}, & \mu_{R} \leq t_{R} \\
 & f_2 = 1, & \mu_{R} > t_{R}
\end{aligned}
\end{equation}
where $f_1(2)$ are defined in Eq. (\ref{eq_f1_f2}) with $t_R = t_L = t, \mu_R = \mu_L = \mu$.

Next, we consider the cases where $\mu_R \neq \mu_L$ or $t_R \neq t_L$. For $\mu_R \neq \mu_L$, $t_R = t_L = t$, we obtain the phase transition condition for $\mu_L \leq t_L$:
\begin{equation}
\begin{aligned}
	&Ve^{|h|} = t,	& \mu_{R} > t_{R}, \\
	&Ve^{|h|} = t_{R},	& \mu_{R} \leq t_{R}.
\end{aligned}
\end{equation}
For $\mu_L > t_L$:
\begin{equation}
\begin{aligned}
	&Ve^{|h|} = \mu, & \mu_{R} > t_{R}, \\
	&Ve^{|h|} = \frac{\sqrt{t^{2} - \mu^{2}}\sqrt{t_{R}^{2} - \mu^{2}} + t t_{R}}{\mu}, & \mu_{R} \leq t_{R}.
\end{aligned}
\end{equation}
We then consider the case where $\mu_R = \mu_L = \mu$, $t_R \neq t_L$. For $\mu_L \leq t_L$:
\begin{equation}
\begin{aligned}
	&Ve^{|h|} = \frac{\mu_{R}\left(t - \sqrt{t^{2} - \mu_{L}^{2}}\right)}{2\mu_{L}} + \frac{\mu_{L}\left(\sqrt{t^{2} - \mu_{L}^{2}} + t\right)}{2\mu_{R}}, & \mu_{R} > t_{R}, \\
	&Ve^{|h|} = \frac{1}{2}\left(\frac{\left(\mu_{R}^{2} - \mu_{L}^{2}\right)\sqrt{t^{2} - \mu_{R}^{2}}}{\mu_{R}^{2}} + \frac{\mu_{L}^{2} t}{\mu_{R}^{2}} + t\right), & \mu_{R} \leq t_{R}.
\end{aligned}
\label{eq_simularity_trtl}
\end{equation}
Notice that, the approximation $\mu_L^2 \approx \mu_L \mu_R$ is used, so the second equation of Eq. (\ref{eq_simularity_trtl}) becomes equivalent to Eq. (\ref{eq_transition_mutL}).

\newpage

\bibliography{NH_offdiagonal_AAH}

\end{document}